\documentclass[prd,twocolumn,amsmath,amssymb,superscriptaddress,floatfix,nofootinbib]{revtex4-2}
\usepackage{txfonts}
\DeclareMathAlphabet{\mathcal}{OMS}{cmsy}{m}{n}
\DeclareSymbolFont{largesymbols}{OMX}{cmex}{m}{n}
\usepackage{amsfonts}
\usepackage[colorlinks=true,citecolor=blue,linkcolor=blue,urlcolor=blue]{hyperref}
\usepackage{color}
\usepackage{graphicx,amsfonts,multirow}
\usepackage{amsmath}
\usepackage{url}
\usepackage{ulem}
\newcommand{\DS}[1]{/\!\!\!#1}

\begin{document}

\title{Probing $D_s^+ \to \eta^{(\prime)} \ell^+\nu_\ell$ semileptonic decay within LCSR under chiral heavy quark effective field theory}

\author{Ruiyu Zhou}
\affiliation{School of Science, Chongqing University of Posts and Telecommunications, Chongqing 400065, P.R. China}

\author{Hai-Bing Fu}
\email{fuhb@gzmu.edu.cn}
\affiliation{Department of Physics, Guizhou Minzu University, Guiyang 550025, P.R. China}
\address{Institute of High Energy Physics, Chinese Academy of Sciences, Beijing 100049, P.R.China}
\author{Yi Zhang}
\affiliation{School of Physics, Beihang University, Beijing 102206, China}
\author{Wei Cheng}
\affiliation{School of Science, Chongqing University of Posts and Telecommunications, Chongqing 400065, P.R. China}
\date{\today}

\begin{abstract}
Motivated by the successful application of Heavy Quark Effective Field Theory in describing decays from heavy to light mesons, this work explores its applicability to the semileptonic decays of charmed mesons. So in this paper we investigate the  $D_s^+\to \eta^{(\prime)} \ell^+ \nu_\ell$ transition form factors using the light-cone sum rules approach within the framework of heavy-quark effective field theory. To address the large uncertainties arsing from the $\eta^{(\prime)}$-meson twist-3 distribution amplitudes, we employ the right-handed chiral correlation function. By applying the  converging simplified series expansion method, we extrapolate the form factors to the entire physical $q^2$-region.  Our analysis yields the branching fractions precise predictions for semi-leptonic decays $D_s^+\to \eta^{(\prime)}\ell^+\nu_\ell$ with : $\mathcal{B}(D_s^+\to\eta \ell^+\nu_\ell)=2.300^{+0.230}_{-0.227}\%$ ($\ell = e$) and $2.249_{-0.206}^{+0.209}\%$ ($\ell = \mu$); $\mathcal{B}(D_s^+\to\eta^\prime \ell^+\nu_\ell)=0.861^{+0.095}_{-0.093}\%$ ($\ell = e$) and $0.821^{+0.082}_{-0.080}\%$ ($\ell = \mu$). The derived lepton flavor universality ratios $R^{\eta}_{\mu,e}=0.977^{+0.008}_{-0.006}$ and $R^{\eta^{\prime}}_{\mu,e} = 0.953^{+0.011}_{-0.009}$ are consistent with lasted BESIII experimental measurements. Additionally, the forward-backward asymmetry parameters $\langle \mathcal{A}^{\eta}_{\rm FB}\rangle=-0.034^{+0.003}_{-0.003}$ and $\langle \mathcal{A}^{\eta^\prime}_{\rm FB}\rangle=-0.073^{+0.007}_{-0.008}$ suggest that no significant violation of lepton flavor universality in this decay process.
\end{abstract}


\maketitle

\textit{Introduction.--}
The study of semileptonic decays in open-charm mesons provides a robust experimental platform for testing the Standard Model (SM) and extracting crucial Cabibbo-Kobayashi-Maskawa (CKM) matrix elements, which are essential for understanding quark mixing and $CP$-violation. These decay processes, particularly those involving heavy-to-light quark transitions (such as $c\to s$), offer unique insights into the interplay between weak and strong interactions in the non-perturbative regime. Recent advancements in collider experiments have enabled precise measurements of specific decay channels such as $D_{s}^{+}\to\eta^{(\prime)}\ell^{+}\nu_{\ell}$ to the forefront of flavor physics research. The presence of strange quarks in the final state $\eta^{(\prime)}$-mesons makes these processes particularly significant. Furthermore, study of the decays $D_s \to \eta^{(\prime)} \ell^+ \nu_\ell$ serves as a valuable probe for elucidating the $\eta-\eta^\prime$ mixing angle and offers a method to explore the complex $\eta-\eta^\prime-$glueball mixing scenario ~\cite{Anisovich:1997dz,DiDonato:2011kr}. These studies are expected to yield profound implications for our understanding of charm physics and broader high-energy phenomenology.

The experimental study of the decay processes involving $D$ mesons containing strange quarks is an ongoing area of research. Over three decades ago, the first measurement of the semi-leptonic decay process $D_s^+ \to \eta$ provided the ratio of the decay branching fraction $\mathcal{B}(D_s^+ \to \eta^{(\prime)} e^+ \nu_e)/\mathcal{B}(D_s^+ \to \phi e^+ \nu_e) =  1.24 \pm 0.12 \pm 0.15 (0.43 \pm 0.11 \pm 0.07)$~\cite{CLEO:1995igy}. Subsequently, the CLEO collaboration utilized data accumulated from the CLEO-c detector to conduct more in-depth analyses of this decay process in 2009~\cite{CLEO:2009dyb} and 2015~\cite{Hietala:2015jqa}. By analyzing data samples of $310~{\rm pb}^{-1}$ and $586~{\rm pb}^{-1}$, they obtained the decay branching fractions as $\mathcal{B}(D_s^+ \to \eta^{(\prime)} e^+ \nu_e) = 2.48 \pm 0.29_{\text{stat}} \pm 0.13_{\text{syst}}\% $($0.91 \pm 0.33_{\text{stat}} \pm 0.05_{\text{syst}}\%$) and $\mathcal{B}(D_s^+ \to \eta^{(\prime)} e^+ \nu_e) = 2.28 \pm 0.14_{\text{stat}} \pm 0.20_{\text{syst}}\%$ ($0.68 \pm 0.15_{\text{stat}} \pm 0.06_{\text{syst}}\%$). In addition, the BESIII collaboration has been continuously making efforts to study this process~\cite{BESIII:2016ult,BESIII:2017ikf,BESIII:2019qci,BESIII:2023ajr,BESIII:2023gbn}.
They have upgraded the reconstruction methods for the final mesons:
for the $\eta$, both the $\gamma\gamma$ and $\pi^0\pi^+\pi^-$ final states are utilized in parallel;
for the $\eta^\prime$, two different final states are employed - either $\eta\pi^+\pi^-$ with $\gamma\rho^0$, or alternatively $\eta\pi^+\pi^-$ with $\gamma\pi^+\pi^-$ for different final lepton flavor.
Owing to the refinement of measurement methods and the continuous accumulation of experimental data, the experiment of this decay process has been significantly enhanced in terms of precision.
Recently, the BESIII Collaboration~\cite{BESIII:2023ajr} performed precise measurements of semi-leptonic decays of $D_s^+ \to \eta^{(\prime)} e^+ \nu_e$ using data with center-of-mass energies of $4.128~\rm{GeV}$ and $4.226~\rm{GeV}$ and integrated luminosity of $7.33~{\rm fb^{-1}}$, resulting in the value of transition form factors (TFFs) at $q^2= 0$, $f_+^{D_s\eta^{(\prime)}}(0) = 0.4642 \pm 0.0073_{\rm stat} \pm 0.0066_{\rm syst} ( 0.540 \pm 0.025_{\rm stat} \pm 0.009_{\rm syst})$ and branching ratios $\mathcal{B}(D_s^+ \to \eta ^{(\prime)} e^+ \nu_e) = 2.255\pm0.039_{\text{stat}}\pm0.051_{\text{syst}}\% (0.810\pm0.038_{\text{stat}}\pm0.024_{\text{syst}}\%)$ and that are more accurate than those obtained from previous experiments~\cite{BESIII:2017ikf}. Shortly thereafter, the BESIII collaboration used the aforementioned electron-positron collision data to perform the first measurement of the semi-leptonic decay dynamics for $D_s^+ \to \eta ^{(\prime) } \mu^+ \nu_{\mu}$ and reported the more precise measurement $f_+^{D_s\eta^{(\prime)}}(0) = 0.465 \pm 0.010_{stat} \pm 0.007_{syst} ( 0.518 \pm 0.038_{stat} \pm 0.012_{syst})$ and $\mathcal{B}(D_s^+ \to \eta ^{(\prime) } \mu^+ \nu_{\mu}) = 2.235 \pm 0.051_{\text{stat}} \pm 0.052_{\text{syst}}\%$ ($0.801 \pm 0.055_{\text{stat}} \pm 0.028_{\text{syst}}\%$)~\cite{BESIII:2023gbn}.
Combining the results of these four decay processes $D_s^+ \to \eta ^{(\prime)} \ell^+ \nu_{\ell}$ with $\ell = (e, \mu)$, the BESIII collaboration tested lepton flavor universality (LFU) in the semi-leptonic decays of these processes, and the findings demonstrate great agreement with the SM predictions. Additionally, the first results on the forward-backward asymmetry of the $\eta^{(\prime)}$-meson were reported, with the measured value of $\langle \mathcal{A}_{\text{FB}}^{\eta^{(\prime)}} \rangle = -0.059 \pm 0.031_{\text{stat}} \pm 0.005_{\text{syst}}$ ($-0.064 \pm 0.079_{\text{stat}} \pm 0.006_{\text{syst}}$).

The increasing precision of experimental measurements in semi-leptonic decays presents new challenges for theoretical studies.
Key observables such as differential decay widths, branching ratios, and CKM matrix elements depend critically on the accurate determination of TFFs.
A considerable gap persists between the TFFs $f^{D_s\eta^{(\prime)}}_{+/0}(q^2)$ derived from some theoretical calculations and the experimental observations~\cite{BESIII:2023gbn}. To probe the strong interaction dynamics in hadronic semi-leptonic decays and address these challenges, theorists employ a variety of non-perturbative methods, including lattice QCD~\cite{Bali:2014pva}, QCD sum rules~\cite{Colangelo:2001cv}, light-cone sum rules (LCSR)~\cite{Duplancic:2015zna, Offen:2013nma, Hu:2021zmy}, and quark models~\cite{Verma:2011yw, Cheng:2017pcq, Wei:2009nc, Melikhov:2000yu, Soni:2018adu, Ivanov:2019nqd}. For examlpe, Lattice QCD studies of the $D_s\to\eta^{(\prime)}$ semi-leptonic decays report the TFFs at maximum recoil point: $f_+^{D_s\eta}(0) = 0.564(11)$ and $f_+^{D_s\eta^\prime}(0) = 0.437(18)$ at $m_{\pi} = 470$ MeV, and $f_+^{D_s\eta}(0) = 0.542(13)$ and $f_+^{D_s\eta^\prime}(0) = 0.404(25)$ at $m_\pi = 370$ MeV~\cite{Bali:2014pva}. Moreover, within the framework of the covariant confining quark model, the author has conducted a detailed investigation of the semi-leptonic decays of $D_{(s)}$ mesons to pseudoscalar or vector meson final states. The results obtained are as follows: $f_+^{D_s\eta}(0) = 0.49\pm0.07$, $f_+^{D_s\eta^\prime}(0) = 0.59\pm0.09$, $B(D_s\to \eta \ell^+ \nu_\ell) = 2.24(2.18)\%$, and $\mathcal{B}(D_s^+\to \eta^\prime \ell^+ \nu_\ell) = 0.83(0.79)\%$, where $\ell =e(\ell=\mu)$, respectively~\cite{Ivanov:2019nqd}.

The heavy quark effective field theory (HQEFT)~\cite{Wu:1992zw, Yan:1999kt, Wu:2000jq, Wang:1999zd, Wang:2000sc, Wang:2000gs} offers a systematic method to differentiate between long-distance and short-distance dynamics, which thereby reducing the complexity of non-perturbative wave functions (WFs) or TFFs and enhances the precision of calculations. The applicability of HQEFT is governed by the criteria $\bar \Lambda / 2 m_{Q} $ and $(v \cdot p) / 2 m_{Q} $ to be smaller than 1, which serve as a criterion for the convergence of the heavy-quark expansion. The effectiveness of the LCSR method within the framework of HQEFT has been well-established in previous studies~\cite{Wang:2005bb, Wu:2006rd, Wang:2010zzi, Zuo:2022ayk, Zuo:2023ksq, Zhou:2019jny, Zuo:2021kui, Zhang:2022opp, Zuo:2024jdf}. Although the applicability of HQEFT to charm systems is less robust than in the bottom sector due to the relatively smaller charm-quark mass, the condition $ m_c \gg \Lambda_{\rm QCD}$ implies that sub-leading power corrections are suppressed to some extent. As a result, the leading-order HQEFT results are still expected to provide a reasonable description of the dominant features of charm-meson decay processes. This study aims to employ this method to perform a systematic analysis of the TFFs for the semileptonic decay process $D_s^+ \to \eta^{(\prime)} \ell^+ \nu_{\ell}$, thereby exploring the effectiveness of the approach in investigating semileptonic decays involving charmed mesons.\\

{\it Theoretical Framework.--}
The transition matrix elements for $D_s^+ \to \eta^{(\prime)} \ell^+ \nu_\ell$ processes can be parameterized through the relevant form factors $f_{+/0}^{D_s \eta^{(\prime)}}(q^2)$ in the following form:
\begin{align}
\langle \eta^{(\prime)}(p)|\bar{s} &\gamma_\mu c |D_s(p+q)\rangle \nonumber\\
=& 2 f_+^{D_s \eta^{(\prime)}}(q^2) p_\mu + \left[f_+^{D_s \eta^{(\prime)}}(q^2)+ f_-^{D_s \eta^{(\prime)}}(q^2) \right] q_\mu \;,  \label{Eq:matrix1}
\end{align}
where
$f_0^{D_s \eta^{(\prime)}}(q^2) = f_+^{D_s \eta^{(\prime)}}(q^2)+ q^2/(m_{D_s}^2-m_{\eta^{(\prime)}}^2) f_-^{D_s \eta^{(\prime)}}(q^2)$,
$p$ and $p+q$ represent the momenta of the final state meson ${\eta^{(\prime)}}$ and the initial state $D_s$-meson, respectively. The $q$ is momentum transfer from the initial meson to lepton pair, and physical range for $q$ is $0 \le q^2 \le (m_{D_s}-m_{\eta^{(\prime)}})^2$. The $D_s$-meson consists of a heavy quark $m_Q$ and a light quark $m_q$, representing a typical heavy-light hadronic system. In the HQEFT framework, these matrix elements can be expressed as ~\cite{Wang:1999zd, Wang:2000sc, Wang:2000gs}:
\begin{align}
\langle \eta^{(\prime)}(p)| \bar{s} \gamma_\mu c |D_s(p+q)\rangle
= - \frac{\sqrt{m_{D_s}}}{\sqrt{\bar\Lambda_{D_s}}}{\rm Tr}[{\eta^{(\prime)}}(v,p){\gamma _\mu }D_{s,v}], \label{Eq:Dseta_HQEFT}
\end{align}
where $\bar \Lambda_{D_s} = m_{D_s} - m_c$, $\eta^{(\prime)} (v,p) = {\gamma ^5}[A(y)+ \DS \hat{p} B(y)]$, $\hat{p}^{\mu}=p^{\mu}/y$, $y=v \cdot p=(m^2_{D_s} + m^2_{\eta^{(\prime)}}-q^2)/2 m_{D_s}$, the effective $D_s$-meson spin function $D_{s,v} = -\sqrt {\bar \Lambda }(1 +\DS v)\gamma^5/2$.
$A(y)$ and $B(y)$ are Lorentz scalar functions.
Based on these theoretical formulations, the TFF $f_\pm(q^2)$ for the semileptonic decays $D_s^+ \to\eta^{(\prime)}\ell^+\nu_\ell$ can be expressed as below:
\begin{eqnarray}
f_\pm^{D_s\eta^{(\prime)}}(q^2) = \frac{\sqrt {\bar\Lambda}}{\sqrt{m_{D_s}\bar\Lambda_{D_s}}}\left[A(y)\pm \frac{m_{D_s}}{y}B(y)\right], \label{Eq:fpm}
\end{eqnarray}
where $\bar \Lambda$ represents the heavy-flavor-independent binding energy, capturing the contribution of the light degrees of freedom within the heavy hadron and $\displaystyle \bar \Lambda = \lim_{m_c \to \infty}  \bar \Lambda_{D_s}$. For the derivation of the sum rules for the two leading order Lorentz scalar functions $A(y)$ and $B(y)$, the following correlation function must be constructed:
\begin{align}
F_\mu (p,q) = i\int d^4x e^{i(q-m_c v) \cdot x}\langle \eta^{(\prime)} (p)|T\{ j_\mu(x),j^\dag_{D_s}(0)\} |0\rangle\;. \label{Eq:correlator}
\end{align}
To mitigate the theoretical uncertainties induced by high-twist distribution amplitudes, this study adopts the chiral current in calculation,
$j_\mu(x) = \bar s(x)\gamma _\mu (1 + \gamma_5)c_v^+(x),
j^{\dag}_{D_s}(0)= i m_c \bar c_v^+(0)  (1 + \gamma _5)s(0),$
where $c^+_v$ is the effective $c$-quark field. Due to the negative parity of the final-state light meson, the final result of the matrix element receives contributions only from pseudoscalar states under the constraint of parity symmetry.

The hadronic representation can be obtained by inserting a complete set of intermediate hadronic states into the correlation function and isolating the pole term of the lowest pseudoscalar state (the $D_s$-meson). Within the HQEFT, the hadronic representation of the correlation function can be expressed as:
\begin{align}
F_\mu^{\rm Had.} (p,q) &= 2 F \frac{A(y)v^\mu  + B(y) \hat p ^\mu }{2 \bar\Lambda_{D_s} - 2(v \cdot k)} \cr
& + \int_{s_0}^\infty  ds \frac{\rho (y,s)}{s - 2(v \cdot k)} + {\rm subtractions}, \label{Eq:Fpq_HQEFT}
\end{align}
where $k$ is the residual momentum of the heavy hadronic, The parameter $F$ is a constant arising from the leading-order term in the heavy-quark expansion of the hadronic matrix element and it serves as an effective decay constant for the meson wihtin this framework. Using the ansatz of the quark-hadron duality the spectral density $\rho (y,s)$ can be obtained~\cite{Shifman:1978bx,Shifman:1978by}.

Subsequently, by employing the operator product expansion (OPE) and $D_s$-meson heavy-quark propagator within HQEFT, the correlator in the deep Euclidean region admits the following representation:
\begin{align}
F_\mu (p,q) &= i\int d^4x e^{i(q-m_b) \cdot x} \int_0^\infty dt \frac{\delta(x-vt)}{2} \nonumber\\
&\times \langle \eta^{(\prime)}(p)|T\{ \bar s(x)\gamma _\mu  \gamma_5 s(0)\} |0\rangle.
\end{align}
By individually applying the Borel transformation to the correlators of the hadronic representation and the OPE, one can effectively diminish the contributions from continuous states in the hadronic representation and high-dimensional condensates in the OPE.
Substituting the LCSR of coefficient functions $A(y)$ and $B(y)$ obtained through the aforementioned steps into Eq.(\ref{Eq:fpm}), we obtain
\begin{align}
&f_\pm^{D_s \eta^{(\prime)}}(q^2) =  - \frac{f_{\eta^{(\prime)}}\sqrt{\bar\Lambda}}{2F\sqrt{m_{D_s}\bar\Lambda_{D_s}}}  \int_0^{s_0^{D_s}} ds e^{\frac{2 \bar \Lambda _{D_s} - s}{T}} \bigg\{\frac{1}{y^2}\frac{\partial }{\partial u}g_2(x) \nonumber\\ &\pm \frac{m_{D_s}}{y} \bigg[ - \phi_{2;\eta^{(\prime)}}(x) + \bigg(\frac{1}{y} \frac{\partial}{\partial u}\bigg)^2 g_1 (x) - \frac{1}{y^2}\frac{\partial }{\partial u}g_2(x)\bigg]\bigg\}\bigg|_{x = 1 - \frac{s}{2y}}\;.  \label{Eq:f+}
\end{align}
where $T$ is the Borel parameter, $s_0^{D_s}$ is the continuum threshold, $\phi_{2;\eta^{(\prime)}}$ is the twist-2 light-cone distribution amplitude (LCDA), and $g_1$ and $g_2$ are twist-4 LCDAs. Before proceeding with the numerical computation of the TFFs, it is necessary to determine key parameters such as the parameters of LCDA, the Borel parameter, and the continuum threshold.\\

\textit{Numerical Analysis.--}
In the calculation of TFFs, several input parameters are required. Specifically, the masses of the $D_s$-meson $m_{D_s} = 1.968{\rm GeV}$, $\eta$-meson $m_{\eta}=0.5478{\rm GeV}$, $\eta^{\prime}$-meson $m_{\eta^{\prime}}=0.9578~{\rm GeV}$, and $c$-quark $m_c=1.27\pm0.02~{\rm GeV}$ are taken from the PDG\cite{ParticleDataGroup:2024cfk}, while the decay constants of $\eta^{(\prime)}$-meson are adopted as $f_{\eta^{(\prime)}} = 0.130(0.157)~{\rm GeV}$, and the leading order $D_s$-meson effective decay constant $F=0.30~{\rm GeV^{3/2}}$~\cite{Wang:2000sc}.

Based on the BHL prescription~\cite{Capri:1983ohx}, the LCHO model's leading-twist WF $\psi_{2;\eta^{(\prime)}}(x, \mathbf{k}_{\perp})$ for the $\eta^{(\prime)}$-meson is derived in the literature \cite{Wu:2011gf}. By considering the relationship between the leading-twist DA and WF of the $\eta^{(\prime)}$-meson, and performing the transverse momentum integration, the following analytical expression for the twist-2 DA was obtained:
\begin{align}
&\phi_{2;\eta^{(\prime)}}(x,\mu) =\frac{\sqrt{3}A_{\eta^{(\prime)}}m_q\beta_{2;\eta^{(\prime)}}\sqrt{x\bar{x}}}{2\sqrt{2}\pi^{3/2}f_{\eta^{(\prime)}}} \nonumber\\
&\qquad\times\left[ 1 + \sum_n B_{n;\eta^{(\prime)}} \times C_n^{3/2} (2x - 1) \right]\nonumber\\
&\qquad\times \left\{ \text{Erf}\left[\sqrt{\frac{m_q^2+\mu^2}{8\beta_{2;\eta^{(\prime)}}^2x\bar{x}}}\right] - \text{Erf}\left[\sqrt{\frac{m_q^2}{8\beta_{2;\eta^{(\prime)}}^2x\bar{x}}}\right] \right\},
\label{phi}
\end{align}
where $\bar{x} = (1 - x)$. $A_{\eta^{(\prime)}}$ is the normalization constant, and $\beta_{2;\eta^{(\prime)}}$ is the harmonious parameter which decided the transverse distribution of WF. Incorporating the two additional parameters $B_{2;\eta^{(\prime)}}$ and $B_{4;\eta^{(\prime)}}$, which characterize the longitudinal profile of the $\eta^{(\prime)}$ -meson distribution amplitude, these four parameters can be determined via the following four constraint conditions. The first one is the normalization condition
$\int_0^1 dx \int d^2 \mathbf{k}_{\perp} \psi_{2;\eta^{(\prime)}}(x, \mathbf{k}_{\perp})/{16\pi^3} = f_{\eta^{(\prime)}}/{\sqrt{24}}$.
\begin{table}[b]
\caption{The $\eta^{(\prime)}$-meson leading-twist LCDA parameters at the initial scale $\mu_0 =1~{\rm GeV}$.}
\begin{tabular}{ccccc}
\hline
\hline
~~~~~~~~~~~~~&~~~~~~$A_{\eta^{(\prime)}}$ ~~~~~~& ~~~~~~$\beta_{2;\eta^{(\prime)}}$~~~~~~& ~~~~~~$B_{2;\eta^{(\prime)}}$~~~~~~& ~~~~~~$B_{4;\eta^{(\prime)}}$~~~~~~ \\
\hline
$\eta$-meson & $11.97^{-0.18}_{+0.24}$ & $4.675^{+0.251}_{-0.345}$ & $-0.117^{+0.018}_{-0.029}$ & $+0.010^{-0.006}_{+0.016}$  \\
$\eta^{\prime}$-meson & $16.42^{-0.57}_{+0.86}$ & $1.822^{-0.156}_{+0.210}$ & $-0.124^{+0.029}_{-0.033}$& $-0.004^{-0.016}_{+0.025}$\\
\hline
\hline
\end{tabular}
\label{DAparameter}
\end{table}
Secondly, the probability of finding quark-antiquark $q\bar{q}$ Fock state in the meson must not exceed unity.
Given that both the $\eta^{(\prime)}$-meson and $\pi$-meson are light pseudo-scalar mesons, we consistently employ the probability parameter $P_{\eta^{(\prime)}}= \int_0^1 dx \int {d^2 \mathbf{k}_{\perp}} |\psi_{2;\eta^{(\prime)}}(x, \mathbf{k}_{\perp})|^2/{16\pi^3} \sim 0.3$ in our subsequent calculations, following the same convention established for the LCWF of pion~\cite{Huang:1994dy}.
The remaining two conditions are specified by the $n$-th order moments $\langle\xi_{2;\eta^{(\prime)}}^n\rangle$ ($n=2, 4$), which incorporate the effects of QCD. The exact values of these moments can be calculated through the background field theory sum rules (BFTSR).
In this study, we determine $B_{2;\eta^{(\prime)}}$, and $B_{4;\eta^{(\prime)}}$ through inverse calculations utilizing the $\langle\xi_{2;\eta^{(\prime)}}^n\rangle$ moments obtained via BFTSR in Ref.~\cite{Hu:2021zmy}. In this paper, we list the parameters of the twist-2 LCDA for the $\eta^{(\prime)}$-meson that we use in Table~\ref{DAparameter}. These parameters correspond to a initial energy scale of $\mu_0 = 1 ~ \text{GeV}$. The typical energy scale in this process is related to the mass of the $D_s$-meson and the $c$-quark, e.g. $\mu \simeq (m_{D_s}^2 - m_c^2)^{1/2}$. For specific calculations, the conventional one-loop evolution equation~\cite{Lepage:1980fj} can be used to evolve to the corresponding energy scale.

\begin{figure}[t]
\begin{center}
\includegraphics[width=0.4\textwidth]{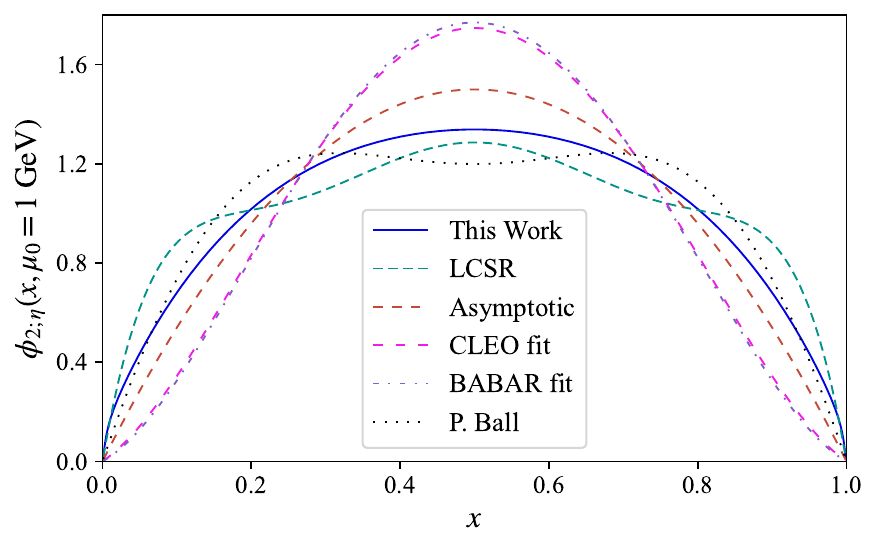}
\end{center}
\caption{The BHL model for the leading twist LCDA $\phi_{2;\eta}(x,\mu_0 = 1~{\rm GeV})$. For comparison, the predictions from the LCSR method \cite{Duplancic:2015zna}, as well as the asymptotic form, the CLEO fit~\cite{CLEO:1997fho}, BABAR fit~\cite{BaBar:2011nrp}, and the results from P. Ball~\cite{Ball:2004ye} are also presented. } \label{DA}
\end{figure}

\begin{table}[b]
\caption{Fitting parameters $b_1$ and $b_2$ for the $D_s^+ \to \eta^{(\prime)}$ TFFs, and quality-of-fit $\Delta$, where the input values consist of both the central values and the upper and lower limits.}
\begin{tabular}{cccc}
\hline
\hline
~~~~~~~~~~~~~~~~~&~~~~~~~~~~~~~~$b_1$~~~~~~~~~~~&~~~~~~~~~~~$b_2$~~~~~~~~~~~&~~~~~~~~~~~~~~$\Delta$~~~~~~~~~~~\\
\hline
$f^{D_s\eta}_+ (q^2)$&$-0.218^{+0.447}_{-0.441}$ & $-2.680^{+3.409}_{-3.139}$ & $0.713^{+0.163}_{-0.275}\%$  \\
$f^{D_s\eta}_0 (q^2)$&$+1.146^{+0.268}_{-0.278}$ & $-5.012^{+1.190}_{-1.035}$ & $0.943^{-0.090}_{-0.011}\%$ \\
$f^{D_s\eta^{\prime}}_+ (q^2)$&$-2.099^{+0.577}_{-0.510}$ & $+198.4^{+22.3}_{-19.5}$ & $0.864^{+0.060}_{-0.057}\%$  \\
$f^{D_s\eta^{\prime}}_0 (q^2)$&$+2.526^{+0.246}_{-0.194}$ & $+8.800^{-1.284}_{+3.932}$ & $0.160^{-0.018}_{+0.027}\%$ \\
\hline
\hline
\end{tabular}
\label{fit}
\end{table}

In Fig.~\ref{DA}, we present the twist-2 DA of the $\eta$-meson within the BHL model, represented by a solid blue line. For comparison, we also show the prediction of LCSR~\cite{Duplancic:2015zna}, the asymptotic form, CLEO collaboration~\cite{CLEO:1997fho}, the BABAR collaboration~\cite{BaBar:2011nrp} and the Ball’s prediction~\cite{Ball:2004ye}. From Fig.~\ref{DA}, it is evident that the dependence of the $\eta$-meson's DA on $x$ remains uncertain. Different models yield distinct behaviors: This single-peak characteristic is further supported by the CLEO~\cite{CLEO:1997fho} fit and BABAR~\cite{BaBar:2011nrp} fit. In striking contrast, the theoretical predictions from LCSR~\cite{Duplancic:2015zna} and the P.Ball's prediction~\cite{Ball:2004ye} consistently demonstrate a distinct double-peaked behavior. The DA within the BHL model initially increases and then decreases with $x$, showing remarkable consistency with the Asymptotic form. The twist-4 LCDAs of two-particle states and the value of these parameters can be found in~\cite{Huang:2001xb,Ball:2006wn}.

\begin{figure}[t]
\includegraphics[width=0.38\textwidth]{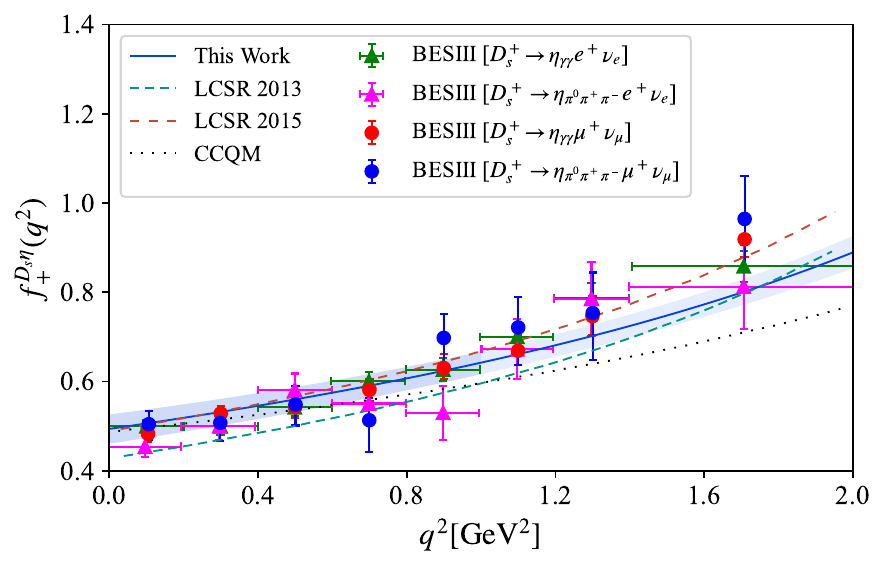}
\includegraphics[width=0.38\textwidth]{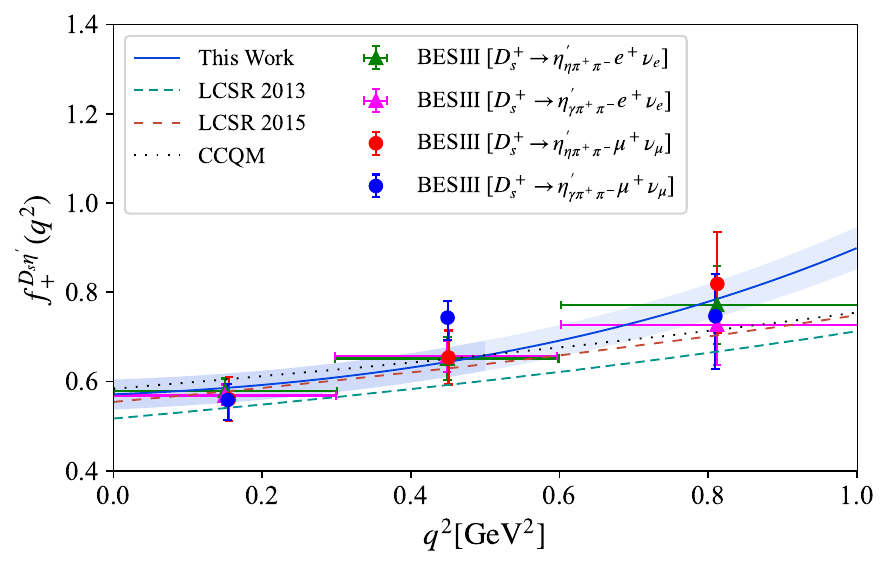}
\caption{The behavior of the $D_s \to \eta^{(\prime)} \ell^+ \nu_l$ TFFs $f_{+}^{D_s \eta^{(\prime)}}(q^2)$ with respect to the transfer momentum $q^2$ is shown in the figure. For comparison, this figure displays theoretical predictions obtained from different approaches (LCSR-2013~\cite{Offen:2013nma}, LCSR-2015~\cite{Duplancic:2015zna}, CCQM~\cite{Ivanov:2019nqd}) along with the most recent experimental measurements by the BESIII Collaboration~\cite{BESIII:2023ajr,BESIII:2023gbn}.}
\label{fp}
\end{figure}

When calculating the TFFs $f^{D_s\eta^{(\prime)}}_{+/0}(q^2)$ for semi-leptonic decays $D_s\to \eta^{(\prime)}$ within the LCSR framework, two crucial parameters must be determined: the continuum threshold $s_0$ and the Borel parameter $T$. Their values should satisfy the following criteria:
(i) The continuum contribution must not exceed $30\%$ of the total LCSR;
(ii) All higher-twist LCDA contributions should remain below $10\%$ of the total LCSR.

The LCSR approach remains reliable only when the energy of the final state meson is not excessively large. In this work, we conservatively select the range $q^2 \in [0, q_{\rm max}^2/2]$ as the applicable domain of the LCSR method.
To obtain the behavior of the TFFs throughout the complete physically allowed kinematic region, we must analytically continue the LCSR's results. In this work, we employ the converging simplified series expansion (SSE) method to extrapolate the obtained TFFs to the full $q^2$ range~\cite{Khodjamirian:2011ub,Bourrely:2008za}:
\begin{align}
f_{+/0}^{D_s \eta^{(\prime)}} (q^2) = \frac{1}{1 - q^2/m_{D_s^*}^2}\bigg[f_{+/0}^{D_s \eta^{(\prime)}} (0)+\sum_{k} b_k z^k(q^2)\bigg]\;, \label{Eq:f+EX}
\end{align}
where the function $z(q^2)$ can be found in Ref.~\cite{Bharucha:2010im}.

\begin{figure}[t]
\begin{center}
\includegraphics[width=0.38\textwidth]{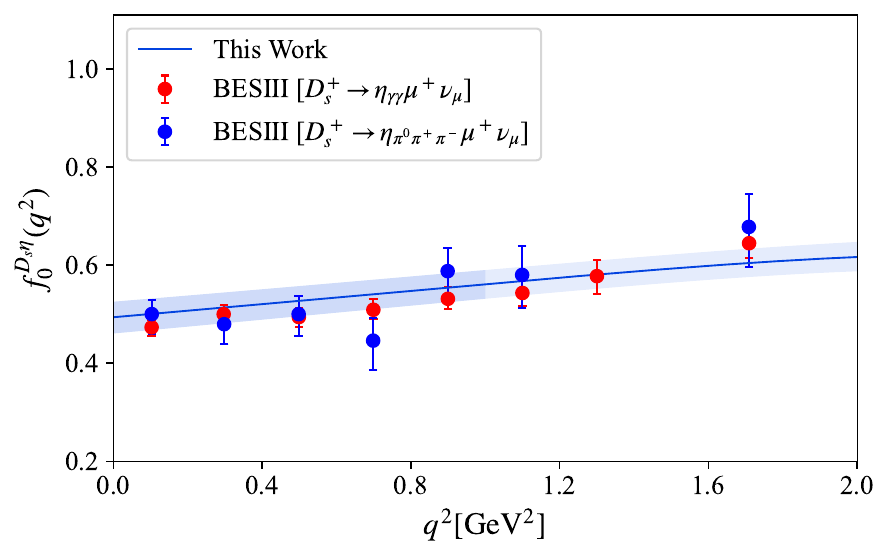}
\includegraphics[width=0.38\textwidth]{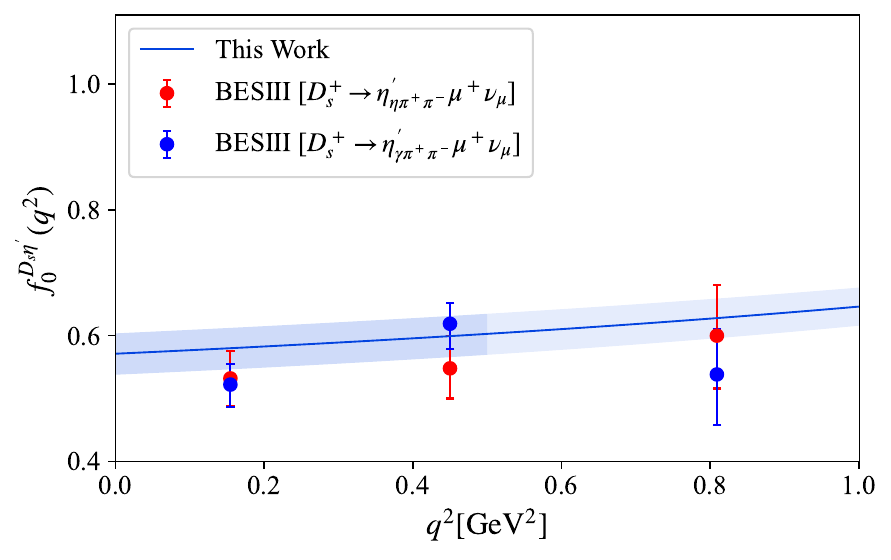}
\end{center}
\caption{The behavior of the TFFs $f_{0}^{D_s \eta^{(\prime)}}(q^2)$ is shown in the figure. The data points with error bars in the figure are consistent with those in Fig.~\ref{fp}.}
\label{f0}
\end{figure}

Then, by fitting the values of the TFFs Eq.~\eqref{Eq:f+} in low and intermediate regions calculated via the LCSRs , the coefficients  $b_1$ and $b_2$ in extrapolation formula Eq.~\eqref{Eq:f+EX} can be determined. The quality-of-fit is defined as: $\Delta  = \sum_{q^2} {|F^{i} - F^{\rm fit}|}/{\sum_{q^2} {|F^{i}|} } \times 100\%$, where ${q^2} \in \{ 0, q_{\rm max}^2/{200},...,100 \times q_{\rm max}^2/{200}\} ~\rm{GeV}^2$. The expansion coefficients $b_i$, listed in Table~\ref{fit}, is performed under the strict requirement that the corresponding quality-of-fit measure $\Delta < 1\%$ to ensure a valid analytic continuation. The resulting $q^2$-evolution of the TFFs are presented in Fig.~\ref{fp} and Fig.~\ref{f0}, with the shaded region systematically combining all theoretical uncertainties mentioned in the preceding sections (e.g. twist-2 LCDA parameters, Borel parameter, continuum threshold, and heavy quark mass). The darker band corresponds to the LCSR predictions of this work, while the lighter band represents the extrapolated results. Meanwhile, this figure also presents predictions from other theoretical approaches and recent experimental results from BESIII. Our results of $f_{+/0}^{D_s \eta}(q^2)$ show good agreement within uncertainties with the latest experimental data, particularly with the $\eta$-meson reconstruction via the di-photon final state. However, the results for $f_{+/0}^{D_s \eta^{\prime}}(q^2)$ demonstrate even better agreement with the experimental data of the three-meson final state $\eta \pi^+ \pi^-$ .

\begin{figure*}[t]
\begin{center}
\includegraphics[width=0.4\textwidth]{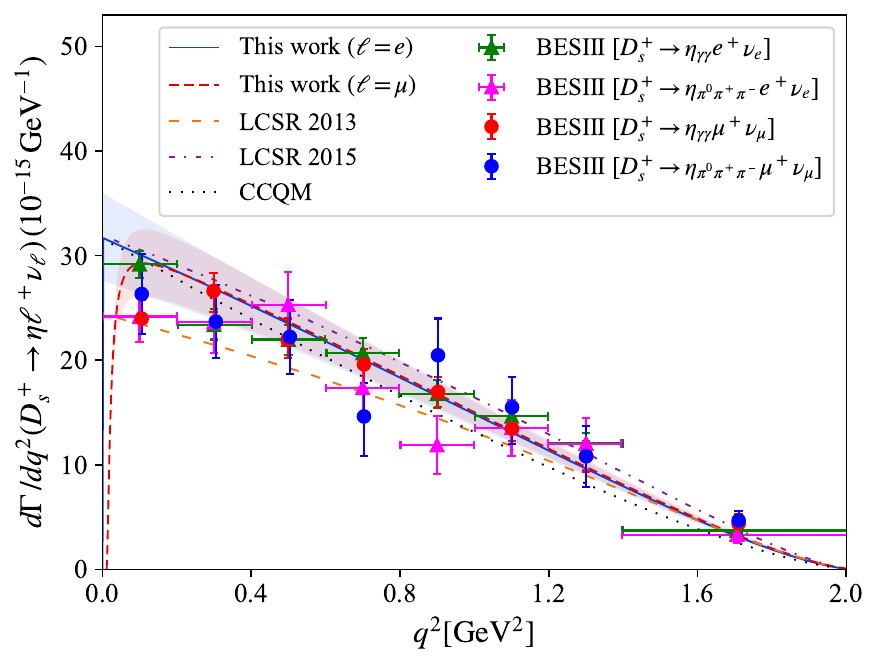}
\includegraphics[width=0.4\textwidth]{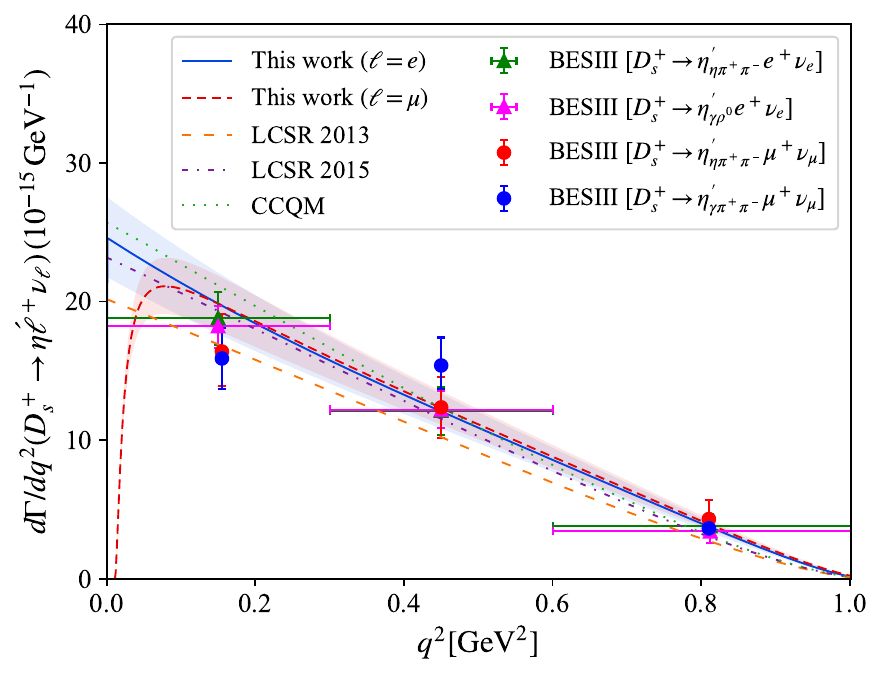}
\end{center}
\caption{The differential decay widths of $D_s \to \eta^{(\prime)} \ell^+ \nu_\ell$, where the uncertainties are squared averages of those from all the mentioned error sources. The other lines and the data points with error bars in the figure are consistent with those in Fig.~\ref{fp}.}
\label{width1}
\end{figure*}
\begin{table*}[t]
\renewcommand\arraystretch{0.9}
\label{BF}
\centering
\caption{Branching factions of $D_s^+\to \eta^{(\prime)} \ell^+ \nu_\ell$ with $\ell =e$ and $\mu$ (in unit $10^{-2}$). The errors are squared averages of all the mentioned error sources. As a comparison, we also present the predictions for various methods.}\label{tab:BF}
\begin{tabular}{l l l l l l}
\hline\hline
Mode~~~~~~~~~~~~~~~~~~~~~~~~~~~~~~   & ${\cal B}(D_s^+ \to\eta e^+\nu_e)$~~~~~~~~~~~~~~~~~~~~~~&${\cal B}(D_s^ + \to\eta \mu^+ \nu_\mu)$~~~~~~~~~~~~~~~~~~~~~~&${\cal B}(D_s^+ \to\eta^{\prime} e^+\nu_e)$~~~~~~~~~~~~~~~~~~~~~~& ${\cal B}(D_s^ + \to\eta^{\prime} \mu^+ \nu_\mu)$
\\\hline
This work (LCSR)                                 & $2.300_{-0.227}^{+0.230}$           & $2.249_{-0.206}^{+0.209}$   &     $0.861_{-0.093}^{+0.095}$  &     $0.821_{-0.080}^{+0.082}$      \\
BESIII-I~\cite{BESIII:2019qci,BESIII:2017ikf}    & $2.323\pm 0.063\pm 0.063$      & $2.42\pm 0.46\pm 0.11$      &$0.824\pm 0.073\pm 0.027$  & $1.06\pm 0.54\pm 0.07$
\\
BESIII-II~\cite{BESIII:2023ajr,BESIII:2023gbn}    & $2.255\pm 0.039\pm 0.051$     & $2.235\pm 0.051\pm 0.052$       & $0.810\pm 0.038\pm 0.024$             & $0.801\pm 0.055\pm 0.028$
\\
CLEO~\cite{Hietala:2015jqa}                      & $2.28\pm 0.14\pm 0.19$              &-    & $0.68\pm 0.15\pm 0.06$                &-                                     \\
CLEO~\cite{CLEO:2009dyb}                        & $2.48\pm 0.29\pm 0.13$              &-    & $0.91\pm 0.33\pm 0.05$                &-                                     \\
PDG~\cite{ParticleDataGroup:2024cfk}                          & $2.26\pm 0.06$                      & $2.4\pm 0.5$  & $0.80\pm 0.04$                        & $1.1\pm 0.5$
\\

LFQM~\cite{Cheng:2017pcq}                        & $2.26\pm 0.21$                      & $2.22\pm 0.20$ &    $0.89\pm 0.09$ &    $0.85\pm 0.08$                     \\
CCQM~\cite{Ivanov:2019nqd}                       & $2.24$                              & $2.18$   & $0.83$ & $0.79$                                \\
LCSR~\cite{Offen:2013nma}                        & $2.00\pm 0.32$                      &-     & $0.75\pm 0.23$                        &-                                    \\
LCSR~\cite{Duplancic:2015zna}                    & $2.40\pm 0.28$                      &-     & $0.79\pm 0.14$                        &-                                    \\
\hline\hline
\end{tabular}
\end{table*}

The differential decay widths of the semileptonic decay $D_s \to \eta^{(\prime)} \ell^+\nu_\ell$ can be written as
\begin{align}
\frac{d\Gamma}{dq^2} &= \frac{G_F^2 |V_{cs}|^2}{24\pi^3} \frac{(q^2 - m_\ell^2)^2 |\boldsymbol p_{\eta^{(\prime)}}|}{q^4 m_{D_s^+}^2}\bigg[ \bigg(1 + \frac{m_\ell^2}{2q^2}\bigg) m_{D_s^+}^2 |p_{\eta^{(\prime)}}|^2 \nonumber\\
&\times |f_+^{\eta^{(\prime)}}(q^2)|^2 + \frac{3m_\ell^2}{8q^2} (m_{D_s^+}^2 - m_{\eta^{(\prime)}}^2)^2 |f_0^{\eta^{(\prime)}}(q^2)|^2 \bigg], \label{Eq:DDF1}
\end{align}
where $|V_{cs}|$ is CKM matrix element, $G_F$ is the Fermi-coupling constant, and we take~\cite{ParticleDataGroup:2024cfk}: $\tau_{D_s}=(0.5012\pm{0.0022}) \times 10^{-12}~s$, $|V_{cs}|=0.975 \pm 0.006 $, $G_F=1.1663787\times 10^{-5}~{ \rm GeV}^{-2}$, $m_e=0.511~{\rm MeV}$ and $m_\mu=105.658~{\rm MeV}$.

To better understand the LFU in the $D_s$-meson, we have calculated the cases where the final charged lepton is an electron or a muon, respectively, and represented them in the figure with pink and blue colors in Fig.~\ref{width1}. The differential widths $d\Gamma(D_s^+\to \eta^{(\prime)} \ell^+ \nu_\ell)/dq^2$ obtained from electron and muon final states show good agreement across the whole physical region, with notable discrepancies only near the maximum recoil point, primarily due to the mass difference between electron $m_e$ and muon $m_\mu$. The figure also presents theoretical predictions from different research groups alongside experimental results from the BESIII collaboration. For the $\eta$-meson decay channel, our results demonstrate excellent consistency with both the di-photon final state and the three-meson final state in both low and high transfer momentum $q^2$-regions, while exhibiting closer alignment with the di-photon characteristics in the intermediate region. Regarding the $\eta^\prime$-meson decay channel, our results maintain overall agreement with both electron and muon final lepton flavor data, though a slight deviation is observed around $q^2 \sim 0.45 \rm{GeV^2}$ when compared to experimental data from the $\eta^\prime \mu^+ \nu_\mu$ final state reconstructed via photon plus $\pi$-meson pairs.

The decay branching fraction ${\cal B}(D_s^+\to \eta^{(\prime)} \ell^+ \nu_\ell)$ can be directly obtained by integrating the differential decay width $d\Gamma (D_s^+\to \eta^{(\prime)} \ell^+ \nu_\ell)/d q^2$. For comparison, Table~\ref{tab:BF} presents our results alongside experimental measurements and other theoretical predictions. Our calculations show good agreement with recent BESIII data ~\cite{BESIII:2023ajr,BESIII:2023gbn} as well as the world average values~\cite{ParticleDataGroup:2024cfk}.

\begin{figure}[t]
\begin{center}
\includegraphics[width=0.4\textwidth]{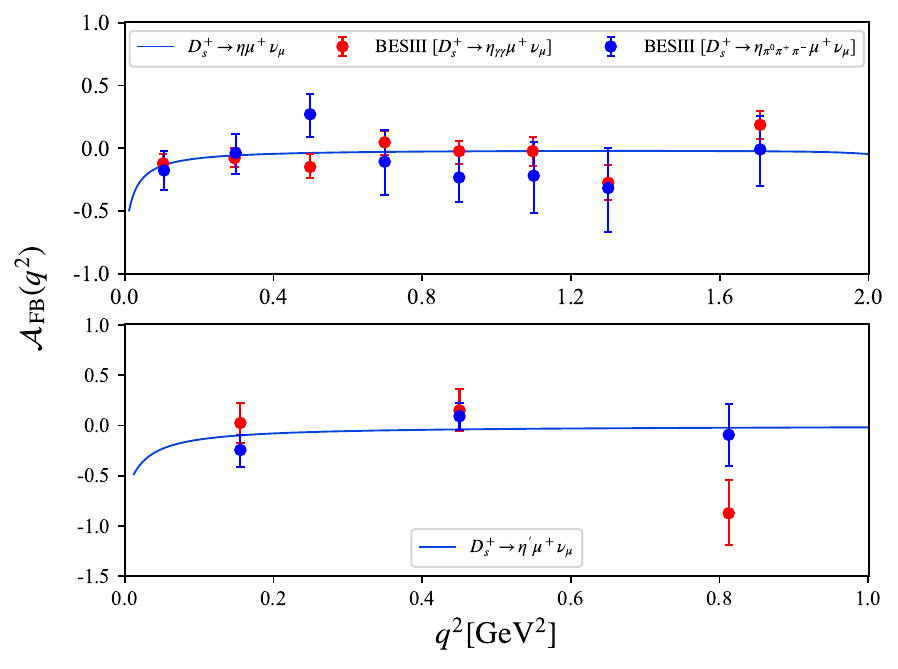}
\end{center}
\caption{The relationship between the forward-backward asymmetry $\mathcal{A}_{\rm FB}(q^2)$ and the square of the momentum transfer $q^2$ in the semi-leptonic decay of $D_s \to \eta^{(\prime)} \mu^+\nu_\mu$ is presented. For comparison, data from the BESIII experimental~\cite{BESIII:2023gbn} are also illustrated.}
\label{AFB}
\end{figure}

The ratio of branching fraction for muon and electron channels, {\it i.e.} $\mathcal{R}_{\mu/e}^{\eta^{(\prime)}} = {\mathcal{B}_{D_s^+ \to \eta^{(\prime)} \mu^+ \nu_\mu}}/{\mathcal{B}_{D_s^+ \to \eta^{(\prime)} e^+ \nu_e}}$ plays a crucial role in probing LFU in $D_s$-meson decays. The current SM prediction for this ratio lies within the range $[0.95, 0.99]$~\cite{Cheng:2017pcq, Ivanov:2019nqd, Hu:2021zmy, BESIII:2023gbn}. Using the decay branching fractions obtained above, our calculations yield $R^{\eta}_{\mu,e}=0.977_{-0.006}^{+0.008}$ and $R^{\eta^{\prime}}_{\mu,e}=0.953_{-0.009}^{+0.011}$, both of which are consistent with the SM expectations.

Based on the definition of the forward-backward asymmetry $\mathcal{A}_{\rm FB}$, we derive the behavior of $D_s$-meson semi-leptonic decays as a function of the momentum transfer $q^2$.
In Fig.~\ref{AFB}, we present the $\mathcal{A}_{\rm FB}(q^2)$ distributions for two decay channels, along with the latest experimental data from the BESIII collaboration for comparison~\cite{BESIII:2023ajr,BESIII:2023gbn}.
The average value of forward-backward asymmetry $\langle \mathcal{A}_{\rm FB} \rangle$ can be obtained $\langle \mathcal{A}^{\eta^{(\prime)}}_{\rm FB} \rangle = -0.034^{+0.003}_{-0.003}(-0.073^{+0.007}_{-0.008})$, which is consistent with the results given by the current BESIII experiment $\langle \mathcal{A}^{\eta^{(\prime)}}_{\rm FB} \rangle_{\rm exp} = -0.059\pm 0.031_{\rm stat}\pm0.005_{\rm syst}(-0.064\pm 0.079_{\rm stat}\pm0.006_{\rm syst})$. This indicates that there is no significant evidence for LFU violation in the semi-leptonic decays of $D_s \to \eta^{(\prime)}$ channel.
\\

\textit{Summary.--}
In this study, we exploit the feature of the $D_s$-meson containing one heavy quark and one light quark, and employ the LCSR within the framework of HQEFT to investigate two TFFs $f_{+/0}^{\eta^{(\prime)}}(q^2)$ of the $D_s$-meson semi-lepton decay channel $D_s^+ \to \eta^{(\prime)} \ell^+ \nu_\ell$. By employing chiral correlation functions, we effectively eliminate theoretical uncertainties introduced by twist-3 distribution amplitudes. Our results for the TFFs at maximum recoil point $f_{+/0}^{\eta^{(\prime)}}(0) =0.494^{+0.032}_{-0.033} (0.571^{+0.032}_{-0.033})$ are consistent with the current experimental measurements within uncertainties.

Employing the SSE method, we present in Fig.~\ref{fp}, Fig.~\ref{f0} and Fig.~\ref{width1} the $q^2$-dependence of the TFFs and differential decay widths across the whole physical region. For the $\eta$-meson, our predictions show better agreement with BESIII experimental data from di-photon final state reconstruction, consistently for both $\ell = e$ and $\ell = \mu$. However, in the case of the $\eta^\prime$-meson the results demonstrate closer consistency with the $\ell = e$ channel measurements. The calculated values of $\mathcal{B}(D_s \to \eta e^+ \nu_e (\mu^+ \nu_\mu) ) = 2.300_{-0.227}^{+0.230}(2.249_{-0.206}^{+0.209})$, $\mathcal{B}(D_s\to \eta^{\prime} e^+ \nu_e (\mu^+ \nu_\mu)) = 0.861_{-0.093}^{+0.095}(0.821_{-0.080}^{+0.082})$ and $R^{\eta^{(\prime)}}_{\mu,e} = 0.977_{-0.006}^{+0.008}(0.953_{-0.009}^{+0.011})$ show excellent consistency with BESIII measurements. To further explore potential LFU violation in $c\to s \ell^+ \nu_\ell $, we analyze the $q^2$-dependent $\mathcal{A}^{\eta^{(\prime)}}_{\rm FB}(q^2)$ in Fig.~\ref{AFB}, obtaining an integrated value of $\langle \mathcal{A}^{\eta^{(\prime)}}_{\rm FB} \rangle = -0.034^{+0.003}_{-0.003}(-0.073^{+0.007}_{-0.008})$. Our findings do not support LFU violation in this decay process. Moreover, as suggested by the above analysis, HQEFT can provide a useful framework for describing the semileptonic decays of charmed mesons. However, given that the charm-quark mass is not sufficiently large, subleading corrections to the form factors $f_+$ and $f_0$
could reach the level of a few tens of percent. A more precise treatment, therefore, requires a systematic consideration of these effects, which will be addressed in our future studies.\\

\textit{Acknowledgments.--}
Hai-Bing Fu would like to thank the IHEP of CAS for their warm and kind hospitality. This work was supported in part by the NSFC under Grant No.12265010, No.12347182 and No.12305109, the Project of Guizhou Provincial Department of Science and Technology under Grants No.MS[2025]219, No.CXTD[2025]030, Chongqing Natural Science Foundation project under Grant No. CSTB2022NSCQ-MSX0534, the Science and Technology Research Project of Chongqing Municipal Education Commission under Grant No.KJQN202300614, and China Postdoctoral Science Foundation Grant under No. 2023M740190.

\bibliography{reference.bib}

\end{document}